\documentclass[a4paper]{article}

\usepackage{INTERSPEECH2019}

\usepackage{makecell}
\usepackage{multirow}
\usepackage{graphicx}
\usepackage{amsmath}
\usepackage{amsxtra}
\usepackage{threeparttable}
\usepackage{subfigure}
\usepackage{stfloats}
\usepackage{tipa}
\usepackage[justification=centering]{caption}

\title{Linguistic-Acoustic Similarity Based Accent Shift for Accent Recognition}
\name{Qijie Shao$^1$, Jinghao Yan$^2$, Jian Kang$^2$, Pengcheng Guo$^1$, Xian Shi$^1$, Pengfei Hu$^2$, Lei Xie$^1$$^{\ast}$ \thanks{*Corresponding author.}}

\address{
  $^1$Audio, Speech, and Language Processing Group (ASLP@NPU), \\
  School of Computer Science, Northwestern Polytechnical University, Xi'an, China \\
  $^2$Tencent Research, Beijing, China}
\email{\{qjshao,xshi\}@npu-aslp.org, pcguo@nwpu-aslp.org, lxie@nwpu.edu.cn\\
\{jinghaoyan,jiankang,alanpfhu\}@tencent.com}

\begin{document}

\maketitle
\begin{abstract}
General accent recognition (AR) models tend to directly extract low-level information from spectrums, which always significantly overfit on speakers or channels. Considering accent can be regarded as a series of shifts relative to native pronunciation, distinguishing accents will be an easier task with accent shift as input. But due to the lack of native utterance as an anchor, estimating the accent shift is difficult. In this paper, we propose linguistic-acoustic similarity based accent shift (LASAS) for AR tasks. For an accent speech utterance, after mapping the corresponding text vector to multiple accent-associated spaces as anchors, its accent shift could be estimated by the similarities between the acoustic embedding and those anchors. Then, we concatenate the accent shift with a dimension-reduced text vector to obtain a linguistic-acoustic bimodal representation. Compared with pure acoustic embedding, the bimodal representation is richer and more clear by taking full advantage of both linguistic and acoustic information, which can effectively improve AR performance. Experiments on Accented English Speech Recognition Challenge (AESRC) dataset show that our method achieves $77.42\%$ accuracy on Test set, obtaining a $6.94\%$ relative improvement over a competitive system in the challenge.
\end{abstract}

\section{Introduction}
Accents are special pronunciations that are generally affected by many factors, e.g., region, native language, and education level~\cite{lippi2012english}. Accent recognition (AR) or accent classification can be used for advertising recommendations and regionally differentiated services. Furthermore, AR is also an important precursor to many speech tasks, such as automatic speech recognition (ASR) and voice assistant. For these tasks, AR has a significant impact on their performance. Therefore, high-performance AR solutions have received extensive attention recently. Accent problems usually include international accents and regional accents. This study mainly focuses on international accents, but our methods can also be used for regional accents.

Some earlier research directly applied x-vector based speaker recognition model ~\cite{snyder2018x} in AR tasks by simply changing the speaker label to accent label~\cite{turan2020achieving,hanani2020spoken,shon2020adi17}. Such models tend to directly extract low-level features (such as the frequency and timbre) from spectrums, resulting in significant overfitting to speakers or channels~\cite{chowdhury2020does}. In recent years, considering that accent is language-related, researchers began to alleviate overfitting by introducing linguistic information, e.g., from ASR. Shi~\textit{et al.}~\cite{shi2021accented} proposed to initialize an AR model with a well-trained ASR encoder and achieved obvious improvement. Many researchers~\cite{zhang2021accent,zhang2021e2e,gao2021end} used multi-task architectures to combine AR and ASR into a union model, which leads the model to focus more on linguistic information. Furthermore, Hamalainen \textit{et al.}~\cite{hamalainen2021finnish} used self-supervised pre-training models~\cite{baevski2020wav2vec,devlin2018bert} to respectively extract linguistic and acoustic embeddings and concatenated them together for an AR task. 

While significant progress has been achieved with the help of linguistic information, the accuracy rates of AR models are still not ideal for downstream tasks. In~\cite{wells1982accents,tjalve2005pronunciation}, researchers showed that accent can be considered as a set of deviations or shifts from standard pronunciation. Therefore, if a native utterance that has the same words as the accent utterance could be obtained, the accent shift could be measured. With this in mind, distinguishing different accents will be an easier task for AR models with accent shift as input. Inspired by this, similar methods~\cite{lee2013mispronunciation,lee2013pronunciation,xiao2018paired} have been applied in pronunciation assessment tasks. In these studies, the pronunciation similarities between the utterance out of the same transcripts from native and non-native speakers are evaluated. Specifically, accent shift was extracted by aligning an accent utterance with a standard pronunciation and comparing their differences. Results showed that accent shift can effectively represent the fine-grained accent characteristics between native and non-native speakers.

However, in AR tasks, it is difficult to obtain such paired utterances which have the same transcripts. In the above methods, native utterance acts as an anchor for accent shift extraction. We consider building a virtual anchor to replace the native utterance in this study. In~\cite{wang2019words,rahman2020integrating,nicolao2015automatic}, word embeddings were regarded as anchors to extract emotion shift from visual and acoustic features. Inspired by these works, in this paper, we propose linguistic-acoustic similarity based accent shift (LASAS) for AR tasks. Specifically, we align the accent speech and its corresponding text by force alignment and map the resulted phoneme-level text sequence into multiple Euclidean spaces. The obtained mapping text vectors are regarded as anchors for accent shift estimation which tries to capture the pronunciation variants of the same word on different accents. Then we use the scale dot-product to calculate the similarities between the acoustic embeddings extracted by a Conformer~\cite{gulati2020conformer} encoder and the mapping text vectors. These similarities show different shift distances and directions for different accents. So their combination could be regarded as an accent shift. Finally, we concatenate the accent shift with a dimension-reduced text vector to obtain a bimodal representation for AR model training. Compared with pure acoustic embedding, the bimodal representation is richer and more clear by taking full advantage of both linguistic and acoustic information, which can effectively improve AR performance. Extensive experiments conducted on the AESRC challenge dataset~\cite{shi2021accented} demonstrate that LASAS is effective and has a clear physical meaning in line with linguistics, which apparently outperforms the direct concatenation of text vector and acoustic embeddings. With ASR-generated text, LASAS achieves $77.42\%$ accuracy on Test set, significantly surpassing a competitive system in the challenge.
\begin{figure*}[htp]
\centering
\vspace{0em}
\includegraphics[width=42.5em]{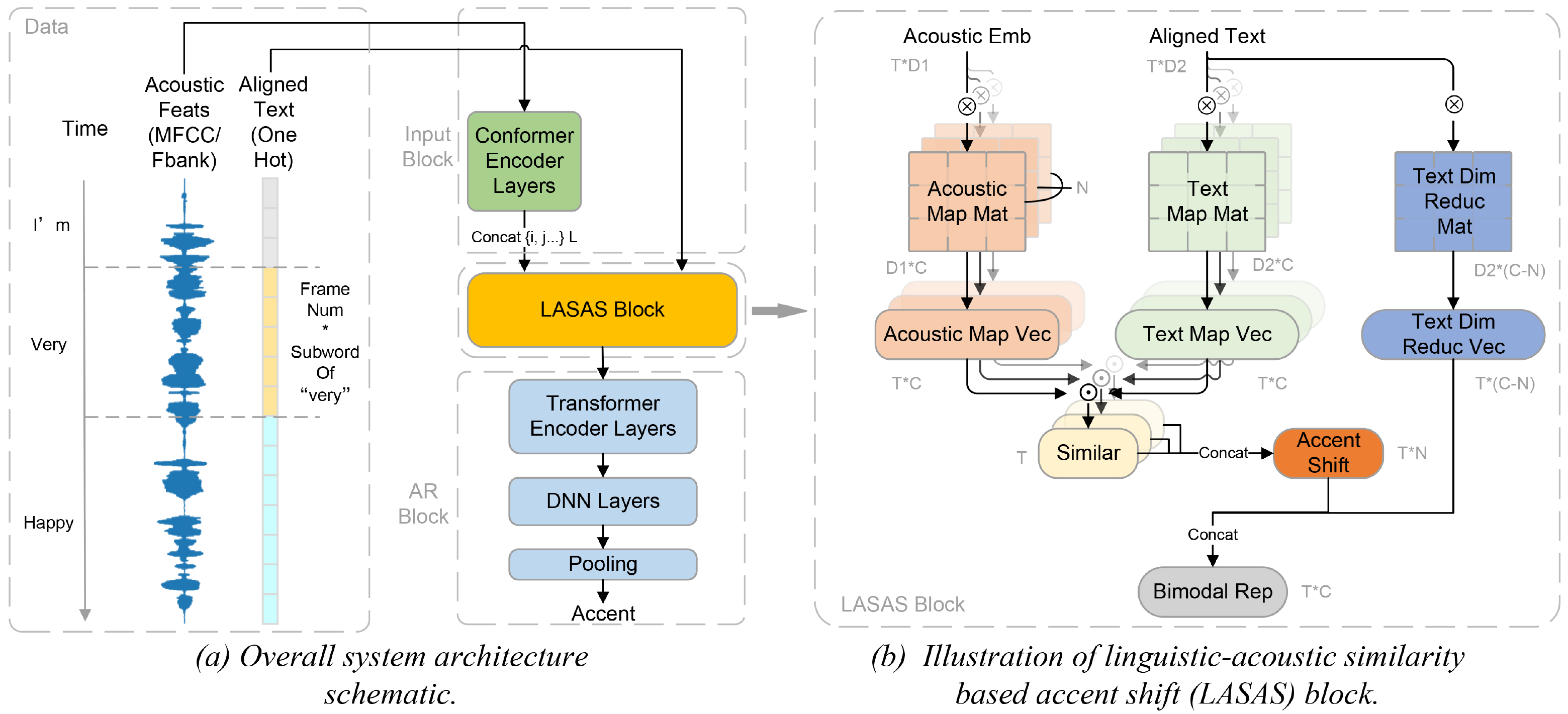}
\vspace{-0.5em}
\caption{The framework of our accent recognition system.}
\vspace{-1em}
\label{fig:system_framework}
\end{figure*}

\section{Method}
\subsection{Overall System Structure}
The purpose of our research is to measure the shift of accent with the help of linguistic information. By incorporating the shift of each accent into the AR training, the model is able to learn more discriminative accent representations over the conventional acoustic features, resulting in better recognition performance. To achieve this goal, we propose a LASAS-based AR model.

As shown in Fig.~\ref{fig:system_framework}~(a), our model consists of three major blocks: (1) \textit{Input block}, which takes acoustic features such as log-Mel filter banks (Fbank) or Mel-frequency cepstral coefficient (MFCC) and aligned text as inputs. We feed the acoustic features into a Conformer~\cite{gulati2020conformer} encoder and then concatenate the outputs of certain encoder layers together as acoustic embeddings. (2) \textit{LASAS block}, which is used to map aligned text and the acoustic embeddings to calculate the similarities as accent shifts, and linguistic-acoustic bimodal representations are obtained. (3) \textit{AR block}, which takes these bimodal representations as input. First, context-sensitive accent information is extracted through a lightweight Transformer~\cite{vaswani2017attention} encoder. Then, the utterance-level accent prediction is obtained after a DNN classifier followed by a statistical pooling layer.

\subsection{Input Block}
Compared with standard pronunciation, accent shift is often manifested in special words or phonemes. Therefore, to evaluate it, pronunciation units need to be given first. In this paper, we use the byte pair encoding (BPE)~\cite{sennrich2015neural} method to obtain subwords, and then take them as pronunciation units.

In ~\cite{lee2013pronunciation}, fine-grained accent features of pronunciation units are extracted by comparing one standard pronunciation with another accented pronunciation. The two sentences are aligned in time before comparison. After the alignment, the accent and standard pronunciation frame at the same time will correspond to the same pronunciation unit. Only in this way can the comparison between two frames be meaningful. As mentioned above, this method needs pairs of accent and standard pronunciation utterances with the same text, which are difficult to obtain in an AR task. So we use aligned text to construct virtual anchors instead of standard pronunciation utterances to measure accent shift. That is, we need an additional ASR system for alignment. The text mentioned here can be manually transcribed or generated by an ASR system, which will be compared in experiments later.

In addition, accents are human voice-related information. Study in~\cite{shi2021accented} shows that the encoder of an ASR model can be used to extract voice-related acoustic embeddings. So we use a Conformer block as an encoder, which is the state-of-the-art architecture in ASR. Finally, several layers' outputs of the encoder are concatenated to form an acoustic embedding.

\subsection{LASAS Block}
As mentioned in~\cite{shi2021accented}, by comparing the differences between a standard pronunciation and another accent utterance, we can extract the physical accent shift. But for AR tasks, this physical shift cannot be obtained under the lack of data pairs. If we build a virtual shift that shows different shift distances and directions on the same pronunciation unit for different accents, such a shift is also meaningful for the subsequent AR model.

To achieve this goal, an anchor is needed, which is related to the speech content and aligned with the speech. We map the aligned text vectors to multiple Euclidean spaces as an anchor set. This method ensures that when the same pronunciation unit appears in utterances with different accents, the anchor is the same. Anchor ${\mathbf{V_{t}^i}}$ is defined as:
\begin{equation}
    \mathbf{V_{t}^i}=\mathbf{X_{t}}\cdot \mathbf{W_{t}^i},\ i\in [1,N]
\end{equation}
where ${N}$ is the number of mapping spaces. ${\mathbf{X_{t}}}$ is an frame-level aligned text vector. ${\mathbf{W_{t}^i}}$ is the text mapping matrix, and ${\mathbf{W_{t}^i}\in\mathbb{R}^{D_{2}\times C}}$. C is the dimension of hidden features, which is shown in Fig.~\ref{fig:system_framework}~(b).

Then, we map an acoustic embedding to the same dimension as the text anchor and calculate the similarity of them by scale dot-product~\cite{vaswani2017attention} frame by frame. This measuring method of similarity for two different dimension vectors is widely used in the attention mechanism~\cite{vaswani2017attention,luong2015effective}. Details are as follows:
\begin{equation}
    \left\{\begin{matrix}
    \mathbf{V_{a}^i}=\mathbf{X_{a}}\cdot \mathbf{W_{a}^i} \\
    S^{i}=\text{DotProd}(\mathbf{V_{a}^i},\ \mathbf{V_{t}^i})/{\sqrt{d_{k}}}
    \end{matrix}\right.
\end{equation}
where ${\mathbf{X_{a}}}$ is acoustic embedding. ${\mathbf{W_{a}^i}}$ is the acoustic mapping matrix, and ${\mathbf{W_{a}^i}\in\mathbb{R}^{D_{1}\times C}}$. And $d_{k}$ is a scale factor, we set $d_{k}=C/N$. ${S^{i}}$ is the similarity value in the ${i^{th}}$ mapping space.

As shown in Fig.~\ref{fig:system_framework}~(b), we use multiple sets of mappings to obtain similarities. All ${\mathbf{W_{a}}}$ and ${\mathbf{W_{t}}}$ are trainable. Therefore, different mappings can measure the accent shifts from different aspects. This method is similar to the multi-head attention mechanism~\cite{vaswani2017attention}. A similarity value can only indicate the proximity between one mapping acoustic vector and one mapping text anchor. However, the combination of multiple similarities can reflect the shift directions and similar degrees of different accents. Therefore, the accent shift ${\mathbf{S}}$ is calculated as:
\begin{equation}
    \mathbf{S}=\text{Concat}(S^{1},\ S^{2},...,\ S^{N})
\end{equation}

Accent shift is a relative representation. For an AR model, it is also necessary to know the reference of this shift. That is, we need to offer the pronunciation unit information corresponding to accent shift. Since each ${\mathbf{V_{t}}}$ mentioned above is related to different accents, the reference should preferably only contain pronunciation unit information. So any ${\mathbf{V_{t}}}$ is not suitable for reference. Therefore, we set up a dimension reduction matrix separately to reduce the dimension of the input text OneHot vector for representing the pure subword information. Then, we directly concatenate the accent shift and the dimension-reduced subword together to form a linguistic-acoustic bimodal representation, which is used as the input for the subsequent AR model. We believe that explicitly preserving the accent shift is better for an AR task than adding it to the subword reference~\cite{wang2019words}. The bimodal representation ${\mathbf{Y_{bm}}}$ is obtained:
\begin{equation}
    \left\{\begin{matrix}
    \mathbf{V_{td}}=\mathbf{X_{t}}\cdot \mathbf{W_{td}}\\
    \mathbf{Y_{bm}}=\text{Concat}(\mathbf{S},\ \mathbf{V_{td}})
    \end{matrix}\right.
\end{equation}

\subsection{Accent Recognition Block}
Since the bimodal representation is frame-level information, we first use a lightweight Transformer encoder with only a few layers to extract context-sensitive accent information. After that, similar to common AR models, DNN is used to reduce the dimension of representations. At last, An utterance-level accent prediction is extracted after a statistical pooling layer.
\section{Experiments}
\subsection{Data And Experimental Settings}
\label{subsec:3-1}
In this study, we use an open-source English accent recognition challenge dataset named AESRC~\cite{shi2021accented} for experiments. Specifically, we use the official Dev and Track1 Test sets for evaluation. There is no speaker overlap in the three sets. The duration of each accent in the training set is 20 hours, while each test set is 2 hours.

As mentioned above, we used an ASR model to align the text, initialize the Conformer encoder, and generate the text in Table~\ref{tab:res_final}. This ASR model is trained by Librispeech (960 hours)~\cite{panayotov2015librispeech} and AESRC data sets with Wenet tools~\cite{zhang2021wenet}. It contains 12 Conformer layers without language model. The CER of this ASR model in the AESRC Dev set is ${6.06\%}$.

In all of our experiments, the Dev set does not participate in training. Our experiments include SpecAugment and model average (top 5), and no other data augmentation. We set the attention dimension of Conformer and Transformer encoders, and the parameter ${C}$ of the LASAS block in Fig.~\ref{fig:system_framework}~(b) to 256. The number of Transformer and DNN layers is 3. And the DNN dimension is halved for each layer. To exclude interference, we use real text in the train and test stage, except ${system 21}$ and ${system 22}$ in Table~\ref{tab:res_final}.

\subsection{Results And Analysis}
\subsubsection{Effectiveness of LASAS}
In Table~\ref{tab:res_effectiveness}, ${system 1}$ is similar to the official baseline in the AESRC challenge with the encoder being replaced from Transformer to Conformer. For further comparison, we remove the LASAS block in Fig.~\ref{fig:system_framework}~(a) and the rest parts are noted as ${system 2}$. ${System 3}$ is a typical LASAS-based AR model and is used as a base model for the following experiments. As shown in Table~\ref{tab:res_effectiveness}, ${system 3}$ is significantly better than ${system 2}$ and ${system 1}$. Without LASAS, even if ${system 2}$ is more refined and complex, its performance is not significantly improved than ${system 1}$. These experiments fully prove that LASAS is effective.
\begin{table}[h]
    \centering
    \footnotesize
    \caption{AR accuracy of LASAS effectiveness experiments.}
    \label{tab:res_effectiveness}
    \resizebox{\linewidth}{!}{
    \begin{tabular}{ccccc}
    \hline
    \textbf{ID} & \textbf{System} & \textbf{Configure} & \textbf{\begin{tabular}[c]{@{}c@{}}Dev\\ (\%)\end{tabular}} & \textbf{\begin{tabular}[c]{@{}c@{}}Test\\ (\%)\end{tabular}} \\ \hline
    1 & Baseline1 & Conformer(12L)+DNN & 73.53 & 68.08 \\ \hline
    2 & Baseline2 & \begin{tabular}[c]{@{}c@{}}Conformer(3 6 9L)\\ +Transformer+DNN\end{tabular} & 75.68 & 67.71 \\ \hline
    3 & LASAS$_{base}$ & \begin{tabular}[c]{@{}c@{}}Conformer(3 6 9L)+LASAS\\ +Transformer+DNN\end{tabular} & \textbf{84.05} & \textbf{75.64} \\ \hline
    \end{tabular}
    }
\end{table}
 
\begin{figure}[h]
    \centering
    \vspace{-0.5em}
    \includegraphics[width=21em]{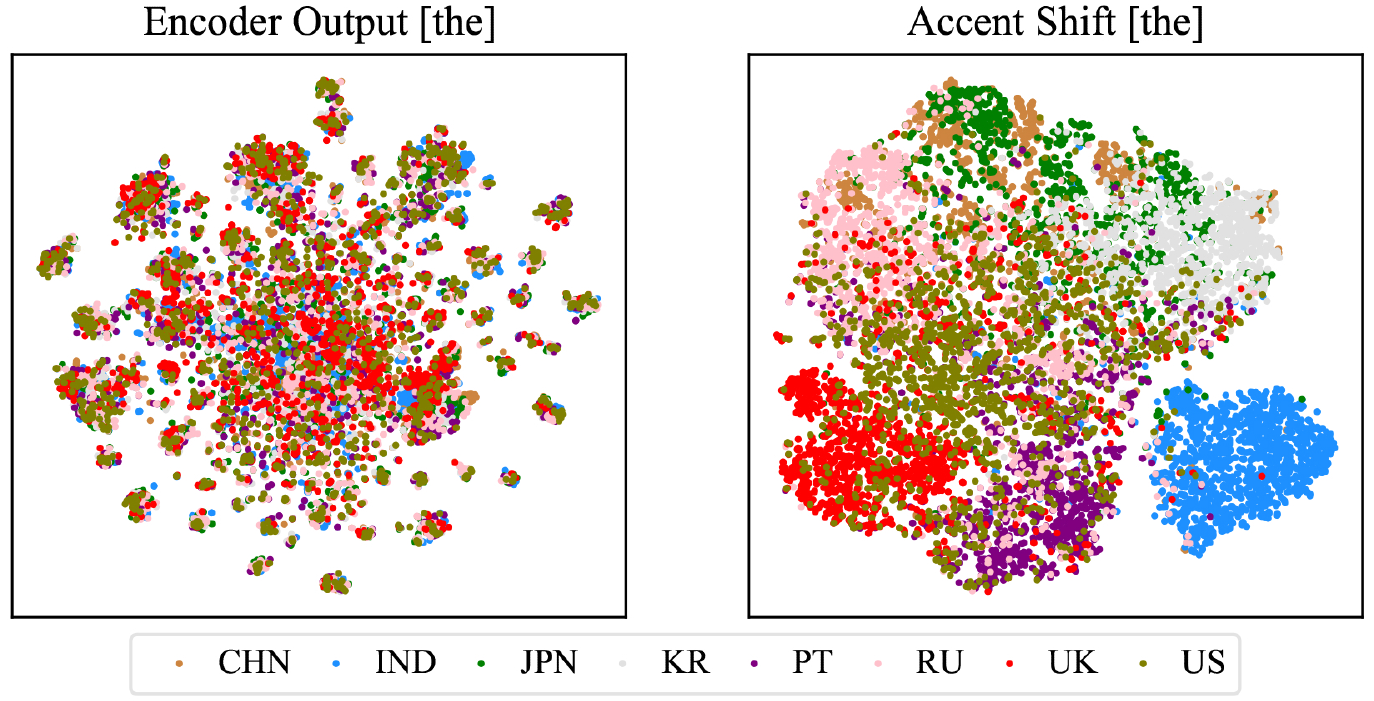}
    \vspace{-0.5em}
    \caption{T-SNE transform of subword ``${the}$''.}
    \vspace{-1em}
    \label{fig:TSNE}
\end{figure}

\begin{figure}[h]
    \centering
    \vspace{-0.5em}
    \includegraphics[width=22em]{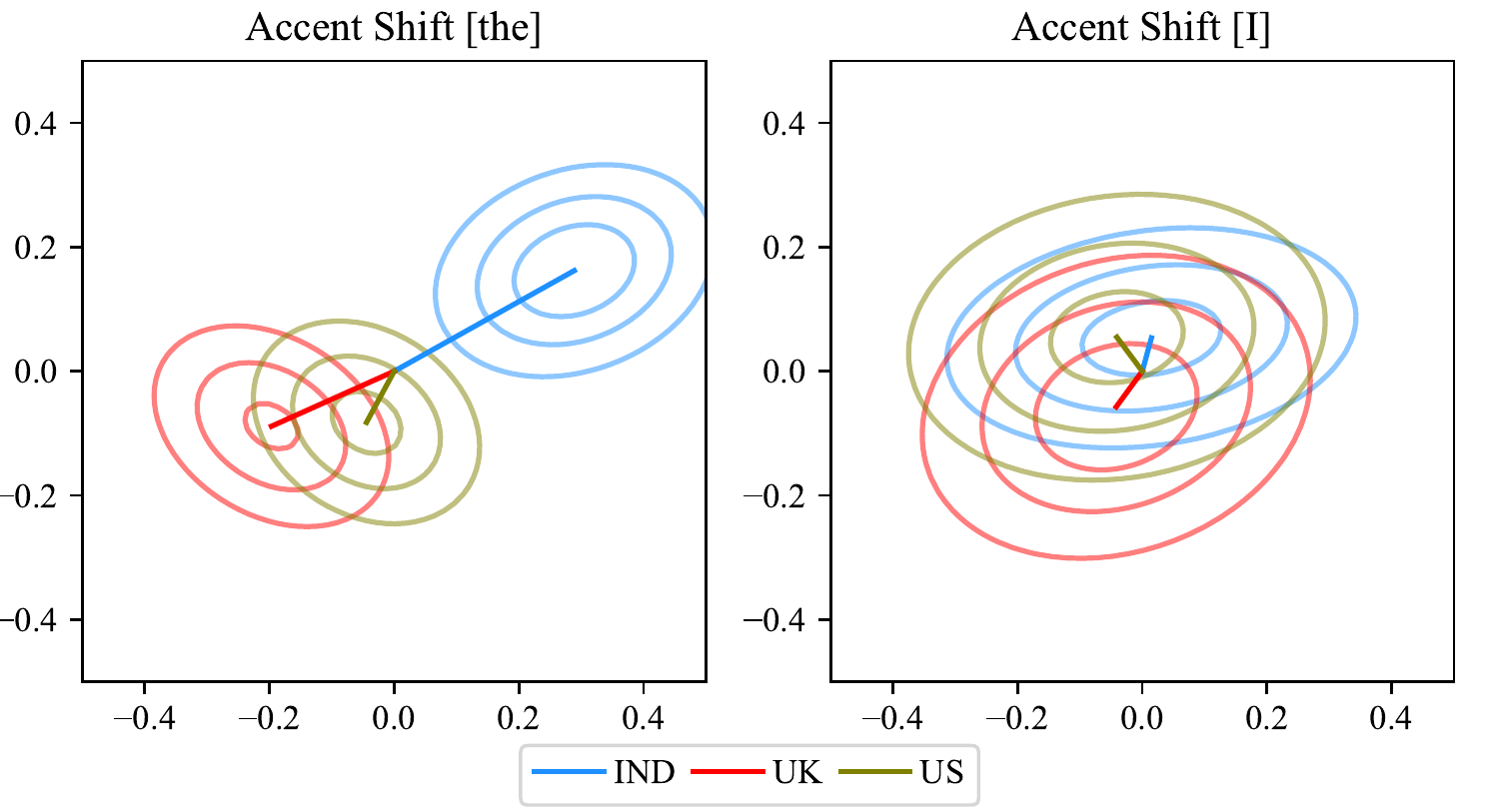}
    \vspace{-0.5em}
    \caption{PCA transform of subword ``${the}$'' and ``${I}$''. The circles are Gaussian contours. The straight lines connect the mean subword centroids of all accents and each individual accent subword centroid, which show accent shift distances.}
    \vspace{-0.5em}
    \label{fig:PCA}
\end{figure}

 We choose subwords ``${the}$'' and ``${I}$'' to visualize accent shift. These two subwords are chosen because, on the one hand, some accents are more significant on ``${the}$'', while most accents have little shift on ``${I}$'', on the other hand, these two words appear more frequently so the results are more credible. We average frame-level accent shifts of each subword segment in the Dev and Test sets. And then project the shifts into 2-dimensional space by t-SNE and PCA transforms, respectively. As shown in Fig.~\ref{fig:TSNE} and Fig.~\ref{fig:PCA}, the distinguishability of accent shift is significantly higher than that of acoustic embedding before LASAS, and the shift distance of ``${the}$'' is significantly larger than ``${I}$''. In particular, for the subword ``${the}$'' in the Indian accent, the clustering effect is prominent in Fig.~\ref{fig:TSNE}, and the shift distance in Fig.~\ref{fig:PCA} is significantly larger than British and American accents. This is because ``[\textipa{D}\textipa{@}]'' is pronounced like ``[d\textipa{@}]'' in the Indian accent, which is unusual in other accent. These phenomena fully prove that the accent shift extracted by LASAS conforms to the laws of linguistics.
 
\subsubsection{Architecture of LASAS}
\label{subsec:3-2-2}
Table~\ref{tab:res_structure} shows the effects of LASAS with different structures. First, comparing ${system 4}$ and ${system 5}$, it makes sense to change the number of mapping spaces. An increment of N can significantly improve AR performance, which indicates that each mapping space learns different accent features. From ${system 6}$, we can see, better results can be obtained with richer acoustic embeddings. Finally, in ${system 7}$, we replace the reference from dimension-reduced text to acoustic embedding. The results of ${system 7}$ show even still includes accent shift, using acoustic embedding as reference is not good as LASAS in ${system 3}$. It shows that using clear and simple linguistic information as a reference will allow the model to learn accent information more easily. 
\begin{table}[h]
    \centering
    \footnotesize
    \caption{AR accuracy of the structure comparison of LASAS.}
    \label{tab:res_structure}
\begin{tabular}{cccccc}
\hline
\textbf{ID} & \textbf{\begin{tabular}[c]{@{}c@{}}Space\\ Num\\ (N)\end{tabular}} & \textbf{\begin{tabular}[c]{@{}c@{}}Encoder\\ Output\\ Layers\end{tabular}} & \textbf{Reference} & \textbf{\begin{tabular}[c]{@{}c@{}}Dev\\ (\%)\end{tabular}} & \textbf{\begin{tabular}[c]{@{}c@{}}Test\\ (\%)\end{tabular}} \\ \hline
3 & 8 & 3,6,9 & Text & 84.05 & 75.64 \\ \hline
4 & 4 & 3,6,9 & Text & 83.46 & 74.22 \\ \hline
5 & 16 & 3,6,9 & Text & 85.42 & 77.16 \\ \hline
6 & 8 & 1$\sim$12 & Text & 84.63 & 78.19 \\ \hline
7 & 8 & 3,6,9 & Acoustic & 81.47 & 73.3 \\ \hline
\end{tabular}
\end{table}

\subsubsection{Comparison of LASAS Versus Direct Concatenation}
\label{subsec:3-2-3}
\begin{figure}[h]
\centering
\vspace{0em}
\includegraphics[width=20em]{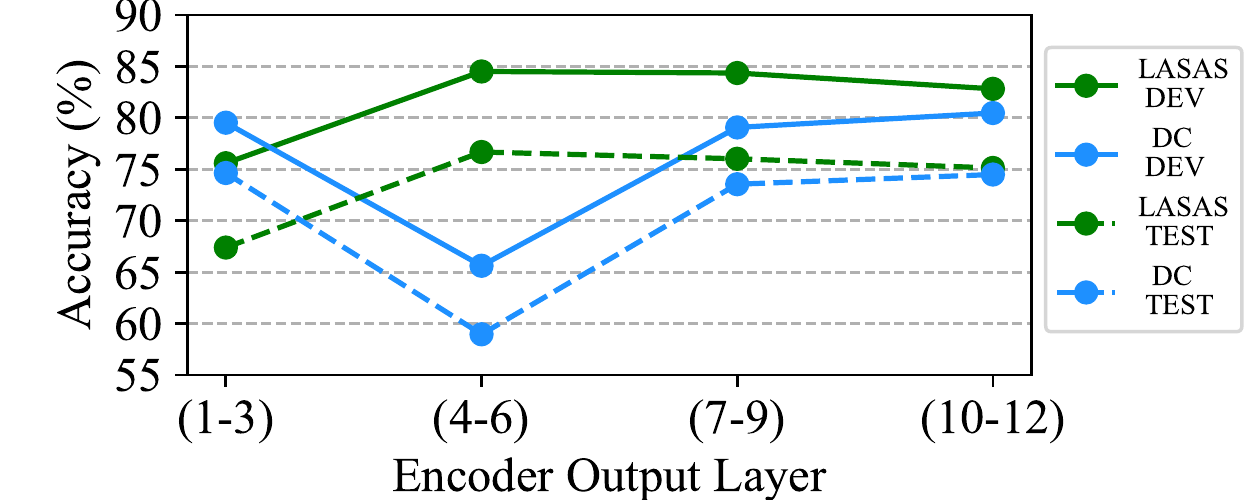}
\vspace{-0.5em}
\caption{LASAS Comparison with direct concatenation (DC) the acoustic embedding and text vector.}
\vspace{-1em}
\label{fig:acoustic_text}
\end{figure}

Considering the introduction of linguistic information may improve AR performance even without LASAS, we conducted comparative experiments (${system 8\sim15}$) by directly concatenating (DC) the text vector and acoustic embedding as the bimodal representation in Fig.~\ref{fig:system_framework}. As shown in Fig.~\ref{fig:acoustic_text}, when using 1-3 layers of encoder output as acoustic embeddings, both methods are less effective. When 4-6 layers were used, LASAS significantly outperforms DC and achieves the best results. After that, with deeper layers are used, the two approaches get closer. This is because the middle layers of the encoder contain more human-voice information, which is required by LASAS. The deeper layers contain too much linguistic information but little related to accents, which leads the performance degeneration when used for LASAS. Overall, LASAS can more fully utilize linguistic-acoustic bimodal information than DC.

\subsubsection{Comparison of Final System Versus AESRC Top-N}
Table~\ref{tab:res_final} shows the results of LASAS and top-level schemes in the AESRC challenge~\cite{shi2021accented}. In ${system 16}$ and ${system 17}$, phone
posteriorgram (PPG) feature extracted from a TDNN-based ASR model is used as inputs for AR models. In addition, ${system 16}$ used model fusion and many kinds of data augmentation methods including TTS. Without model fusion and comprehensive data augmentation (except SpecAug), our schemes ${system 20\sim22}$ achieve similar performance with ${system 17}$ in the Dev set. In ${system 18}$, researchers trained a hybrid CTC/attention-based ASR model and made the model learned to predict text and accent at the same time by transfer learning. In ${system 19}$, an ASR-AR multitask architecture is used. And the ASR-generated features are used for the AR task, which contained sufficient implicit linguistic information. Our schemes ${system 20\sim22}$ are significantly surpassing ${system 18}$ and ${system 19}$, which fully proves the advantages of LASAS.
\begin{table}[h]
    \centering
    \footnotesize
    \caption{AR accuracy of LASAS and AESRC top-level schemes. Real and ASR in the table represent using real text or ASR-generated text, respectively. Details of the ASR model are mentioned in Section~\ref{subsec:3-1}.}
    \label{tab:res_final}
    \begin{tabular}{ccccc}
    \hline
    \textbf{ID} & \multicolumn{2}{c}{\textbf{System}} & \textbf{\begin{tabular}[c]{@{}c@{}}Dev\\ (\%)\end{tabular}} & \textbf{\begin{tabular}[c]{@{}c@{}}Test\\ (\%)\end{tabular}} \\ \hline
    16 & \multicolumn{2}{c}{Top1: AR (PPG) + Data Aug (TTS)} & 91.3 & 83.63 \\ \hline
    17 & \multicolumn{2}{c}{Top1 w/o Data Aug} & 84.51 & - \\ \hline
    18 & \multicolumn{2}{c}{Top2: AR+ASR Fusion} & 80.98 & 72.39 \\ \hline
    19 & \multicolumn{2}{c}{Top3: AR+ASR Multitask} & 81.1 & 69.63 \\ \hline
    20 & \multirow{3}{*}{\textbf{\begin{tabular}[c]{@{}c@{}}LASAS$_{final}$:\\ N=16,\\ Enc=(4$\sim$9)L\end{tabular}}} & Real$_{train}$+Real$_{test}$ & 85.39 & 77.79 \\ \cline{1-1} \cline{3-5} 
    21 &  & Real$_{train}$+ASR$_{test}$ & 85.12 & 77.18 \\ \cline{1-1} \cline{3-5} 
    \textbf{22} &  & \textbf{ASR$_{train}$+ASR$_{test}$} & \textbf{84.88} & \textbf{77.42} \\ \hline
    \end{tabular}
\end{table}

In ${system 20\sim22}$, we further optimize the model according to the conclusions in Section~\ref{subsec:3-2-2} and Section~\ref{subsec:3-2-3} by using 4-9 layers' outputs of the Conformer encoder and setting the number of mapping spaces to 16. Then, we investigate the impact of text reliability on LASAS. As expected, in ${system 20}$, LASAS achieves the best results when using manually annotated transcripts (real text). If using real text for training, and ASR-generated text in the test stage, the accuracy of the Dev and Test sets drops by ${0.32\%}$ and ${0.78\%}$, respectively. If in the training stage we also use ASR-generated text, this gap may be alleviated, and the respective accuracy of ${system 22}$ on the Test set drops by only ${0.48\%}$. This means that the LASAS has high robustness to text errors. Since LASAS integrates the information of the whole utterance, it can maintain good performance even if some characters are wrong. At last, ${system 22}$ achieves $84.88\%$ and $77.42\%$ accuracy on the Dev and Test set, obtaining a $4.81\%$ and $6.94\%$ relative improvement over ${system 18}$.
\section{Conclusions}
In this paper, we propose a LASAS-based AR model. We extract accent shift by LASAS block and concatenate it with a text reference to form a bimodal representation for AR tasks. Experiments show that our method effectively improve AR performance. We visualize the accent shift and show its rationality. Furthermore, our scheme significantly outperforms the method that directly concatenates acoustic embedding and text vector, and also shows superior performance in the AESRC challenge dataset. In the future, we hope to improve LASAS for accent ASR tasks or text-dependent speaker verification tasks.

\vfill\pagebreak
\bibliographystyle{IEEEtran}
\bibliography{6_Ref}

\end{document}